\documentclass[%
 reprint,
 amsmath,amssymb,
 aps,
]{revtex4-1}

\usepackage{graphicx}% Include figure files
\usepackage{dcolumn}% Align table columns on decimal point
\usepackage{bm}% bold math
\begin{document}

\preprint{APS/123-QED}

\title{Electron Effective Mass in Graphene}% Force line breaks with \\
% \thanks{Thanks to:  Prof. Shlomo Ruschin of Tel Aviv University, and Dr. Leon Altschul of SDL Ltd. for reviewing this work.}%
\author{Viktor Ariel and Amir Natan}
\affiliation{%
 Department of Physical Electronics\\
 Tel-Aviv University
}%

\begin{abstract}
The particle effective mass in graphene is a challenging concept because the commonly used theoretical expression is mathematically divergent. In this paper, we use basic principles to present a simple theoretical expression for the effective mass that is suitable for both parabolic and non-parabolic isotropic  materials. We demonstrate  that this definition is consistent with the definition of the cyclotron effective mass, which is one of the common methods for effective mass measurement in solid state materials. We apply the proposed theoretical definition to graphene and demonstrate  linear dependence of the effective mass on momentum, as confirmed by experimental cyclotron resonance measurements. Therefore,  the proposed definition of the effective mass can be used for non-parabolic materials such as graphene. \\
\end{abstract}

\pacs{Valid PACS appear here}% PACS, the Physics and Astronomy
\maketitle

%\tableofcontents

\section{\label{sec:level1}Introduction\protect\\}
 In solid state materials, many different effective mass definitions are used such as  the conductivity, density-of-states, optical, and cyclotron effective mass \cite{Ashcroft}, \cite{Seeger},  \cite{Harrison}, \cite{Kittel}. It can be shown that the  most commonly used theoretical definition of the effective mass, using the second derivative of the energy, is limited to parabolic energy dispersion.  An alternative definition should be used for non-parabolic materials such as graphene \cite{Zawadzki}, \cite{ArielPaper}. 

In a recent work \cite{ArielArxiv},  a  theoretical expression for the effective mass was presented that seems suitable for  both parabolic  and non-parabolic materials. In this work,  we demonstrate that this definition of the effective mass is consistent with the definition of the cyclotron mass, which is commonly used for experimental measurements of the effective mass.  Finally, we apply this definition to graphene and show that it is in agreement with the experimentally observed linear dependence between the cyclotron mass and  momentum.

\section{Definition of the Effective Mass }

We use the traditional semi-classical approach by associating  particles with  wave-packets. This is a standard approach used in  solid state physics for calculations of the energy-band structure and charge transport properties. 

Applying wave-particle duality, we associate particle velocity with the group velocity of the wave-packet and  the particle momentum with the crystal momentum $p=\hbar k$.  Then,
\begin{equation}
 v=v_g \simeq  \frac {1} {\hbar} \frac {\partial E} { \partial  k} \ .
\label{eq:group}
\end{equation}

The effective mass appears as a proportionality factor between the particle momentum and the group velocity of the wave-packet 
\begin{equation}
 p =  \hbar k \simeq  m^* v_g \ .
\label{eq:momentum}
\end{equation}

Based on equations  (\ref{eq:group})  and   (\ref{eq:momentum})   we immediately obtain 

\begin{equation}
 m^*(E,k) =\frac {p} {v_g} =   \hbar^2 k \left ( \frac {\partial E} { \partial  k} \right)^{-1}  \ .
\label{eq:mass}
\end{equation}

Note that the  effective mass defined above (\ref{eq:mass}) is generally energy and momentum dependent. This theoretical definition of the effective mass  was previously developed \cite{Zawadzki},  \cite{ArielArxiv} and   sometimes reffered to as the optical effective mass \cite{Seeger}. Often, the following theoretical definition of the effective mass is  used  in solid state physics \cite{Kittel}:
\begin{equation}
m^*=\frac {1} {\partial^2 E/ \partial k^2} \ .
\label{eq:mass-parab}
\end{equation}
This expression is derived using an implicit assumption of the parabolic $E(k)$ relationship  and therefore should not be applied to non-parabolic $E(k)$, 
as was demonstrated in \cite{ArielArxiv}. On the other hand, the effective mass defined by Eq.\ (\ref{eq:mass}) is not 
based on such an assumption and should be suitable for an arbitrary $E(k)$ relationship including non-parabolic solid state materials such as graphene. Furthermore, for the case of a parabolic material, this new definition gives exactly the same result as the traditional Eq.\ (\ref{eq:mass-parab}).

\section{Comparison of the cyclotron mass to the effective mass}

In the presence of a constant magnetic field, the part of $k$ parallel to the field, $k_{||}$, is a constant of motion. The part of $k$ perpenidcular to the field, $k_\bot$, changes in time so that the elctron moves in k-space in a closed line of constant energy. The electron therefore follows a spiral path whose projection on a plane perpendicular to the field is a closed path in both real space and k-space (See chapter 12 in \cite{Ashcroft}). Similar to the free electron case, the cyclotron resonance frequency for an electron in a solid is defined using the concept of the cycltron effective mass:

\begin{equation}
\omega_c = \frac { e H} {m^* c}\ .
\label{eq:solid_electron_cyclotron}
\end{equation}
This leads to the cyclotron effective mass, $m^*$, given as in \cite{Ashcroft}:

\begin{equation}
m^*=\frac{\hbar^2}{2\pi}\left (\frac{\partial A(E)}{\partial E}\right)
\label{eq:cyclotron_area_mass}
\end{equation}

Where $A(E)$ is the area bound by the closed path in $k_\bot$ space that is traveled by the electron at the given energy level $E$.
In the absense of scattering, the particle trajectory in an isotropic material is a circle perpendicular to the direction of the field and so:
 
\begin{equation}
 A(k)= \pi k^2 .
\label{eq:Energy-surface}
\end{equation}

For such an isotropic material we can apply the chain rule in Eq. (\ref{eq:cyclotron_area_mass}) and write:

\begin{equation}
m^*=\frac{\hbar^2}{2\pi}\frac{\partial A(k)}{\partial k}\frac{\partial k}{\partial E}=\hbar^2 k \left(\frac{\partial E}{\partial k}\right)^{-1}
\end{equation}

This demonstrates that the cyclotron effective mass \cite{Ashcroft} for an isotropic material is equivalent to the theoretical definition of the effective mass in  Eq. (\ref{eq:mass}). This result would stay correct for any dispersion relation $E(k)$ as long as the material is isotropic. 

It is also possible to derive Eq.\ (\ref{eq:solid_electron_cyclotron}) directly from Eq.~(\ref{eq:mass}).  We write the equation of motion for $k_\bot$ in an isotropic material:

\begin{equation}
\hbar \frac{\partial \overrightarrow{k_\bot}}{\partial t}=-\frac{e}{c} \overrightarrow{v_g} \times \overrightarrow{H}
\end{equation}

Since $k_\bot$ and $v_g$ are perpendicular to $H$, we can use the group velocity  Eq. (\ref{eq:momentum}) to  write:

\begin{equation}
\hbar \omega_c k = \frac{e \hbar k H}{m^* c} \Rightarrow \omega_c = \frac{e H}{m^* c}
\end{equation}

Thus, we got the cyclotron resonance directly from the semi-classical equations of motion and from the effective mass as defined by Eq. (\ref{eq:mass}).

\section{Application to graphene}

The two-dimensional electron gas observed in graphene can be described by the following isotropic $E(k)$ relationship in the vicinity of the Dirac points \cite{Castro}:  

\begin{equation}
E \simeq  \hbar k v_f \ ,
\label{eq:graphene-dispersion}
\end{equation}
where $v_f$ is the Fermi velocity. 
Using this dispersion relationship (\ref{eq:graphene-dispersion}) and the definition of the group velocity (\ref{eq:group}) leads to

\begin{equation}
v_g  \simeq  {v_f}  .
\label{eq:graphene-group}
\end{equation}

The definition of the effective mass (\ref{eq:mass}) applied to graphene dispersion relationship  (\ref{eq:graphene-dispersion}) results in
\begin{equation}
 m^*  \simeq  \frac {p} {v_g}\simeq \frac {\hbar k} {v_f}  .
\label{eq:graphen-mass}
\end{equation}

The above linear dependence of the particle effective mass on momentum  in graphene was  confirmed by cyclotron resonance measurements  \cite{Castro}. 
Note that in graphene  the parabolic effective mass definition (\ref{eq:mass-parab})  leads to a mathematically divergent expression.   

The proposed definition of the effective mass  Eq.\ (\ref{eq:mass}) can be successfully  
applied  to  graphene and can be experimentally verified.

\section{Conclusions}

In this work, we demonstrated that a simple  theoretical definition of the effective mass (\ref{eq:mass}) is compatible with the  definition of the  cyclotron effective mass, which is a common method for experimental measurement of the carrier effective mass in solid state materials.  We  show that in graphene this theoretical  definition of the effective mass results in correct experimentally observed dependence of  the  carrier effective mass on momentum.  Therefore, it appears that the proposed definition of effective mass is suitable  for theoretical and experimental studies of particle transport properties in non-parabolic solid state materials such as graphene.


\begin{thebibliography}{9}



\bibitem{Ashcroft}
N. W. Aschcroft, 
  \emph{Solid State Theory}
(Holt, Rinehart, and Wilson,  1976).

\bibitem{Seeger}
K. Seeger, 
\emph{Semiconductor Physics}
(Springer-Verlag, 1985).

\bibitem{Harrison}
W. A. Harrison, 
  \emph{Solid State Theory}
(Dover,  1979).


\bibitem{Kittel}
C. Kittel, 
\emph{Introduction to Solid State Physics}
(Wiley, 2nd ed., 1986).



\bibitem{Zawadzki}
W. Zawadzki, S. Klahn, U. Merkt, Phys. Rev. Lett.  {\bf 55}, 983 (1985).

\bibitem{ArielPaper}
V. Ariel, A. Fraenkel, E. Finkman, J. Appl. Phys. {\bf 71}, 4382 (1992).


\bibitem{ArielArxiv}
V. Ariel, arXiv:1205.3995v1 [physics.gen-ph], (2012).

\bibitem{Shockley}
W. Shockley, Phys. Rev., {\bf 79}, 191 (1950).

\bibitem{Castro}
A. H. Castro Neto, F. Guinea,  N. M. R. Peres, K. S. Novoselov, A. K. Geim,
“The electronic properties of graphene”, Rev.  Mod. Phys., {\bf 81}, 109 (2009).



\end{thebibliography}
\end{document}